\documentclass[preprint]{aastex631}

\newcommand{\FRB}{FRB~20190520B}

\begin{document}

\title{Simultaneous multi-wavelength observations of the repeating fast radio burst FRB~20190520B with Swift and FAST}

\correspondingauthor{Wenfei Yu}
\email{wenfei@shao.ac.cn}

\author[0000-0002-5385-9586]{Zhen Yan}
\affiliation{Shanghai Astronomical Observatory, Chinese Academy of Sciences, 80 Nandan Road,
Shanghai 200030, China}

\author[0000-0002-3844-9677]{Wenfei Yu}
\affiliation{Shanghai Astronomical Observatory, Chinese Academy of Sciences, 80 Nandan Road,
Shanghai 200030, China}

\author[0000-0001-5624-2613]{Kim L. Page}
\affiliation{School of Physics \& Astronomy, University of Leicester, LE1 7RH, UK}

\author{Jie Lin}
\affiliation{CAS Key laboratory for Research in Galaxies and Cosmology, Department of Astronomy, University of Science and Technology of China, Hefei 230026, China}
\affiliation{School of Astronomy and Space Sciences, University of Science and Technology of China, Hefei 230026, China}

\author{Di Li}
\affiliation{Department of Astronomy, Tsinghua University, Beijing 100084, China}
\affiliation{National Astronomical Observatories, Chinese Academy of Sciences, A20 Datun Road, Chaoyang District, Beijing 100101, China}
\affiliation{Zhejiang Lab, Hangzhou, Zhejiang 311121, People’s Republic of China}

\author{Chenhui Niu}
\affiliation{National Astronomical Observatories, Chinese Academy of Sciences, Beijing 100012, China}
\affiliation{Institute of Astrophysics, Central China Normal University, Wuhan 430079, China}

\author{Casey Law}
\affiliation{Cahill Center for Astronomy and Astrophysics, MC 249-17 California Institute of Technology, Pasadena, CA 91125, USA}
\affiliation{Owens Valley Radio Observatory, California Institute of Technology, 100 Leighton Lane, Big Pine, CA, 93513, USA}

\author[0000-0002-9725-2524]{Bing Zhang}
\affiliation{The Nevada Center for Astrophysics, University of Nevada, Las Vegas, Las Vegas, NV 89154, USA}
\affiliation{Department of Physics and Astronomy, University of Nevada, Las Vegas, Las Vegas, NV 89154, USA}

\author{Shami Chatterjee}
\affiliation{Cornell Center for Astrophysics and Planetary Science, and Department of Astronomy, Cornell University, Ithaca, NY, USA.}

\author[0000-0002-8086-4049]{Xian Zhang}
\affiliation{Shanghai Astronomical Observatory, Chinese Academy of Sciences, 80 Nandan Road, Shanghai 200030, China}
\affiliation{University of Chinese Academy of Sciences, 19A Yuquanlu, Beijing 100049, China}

\author[0000-0001-8057-0633]{Reshma Anna-Thomas}
\affiliation{ West Virginia University, Department of Physics and Astronomy, P.O. Box 6315, Morgantown, WV, USA}
\affiliation{Center for Gravitational Waves and Cosmology, West Virginia University, Chestnut Ridge Research Building, Morgantown, WV, USA}

\begin{abstract}
Among several dozen known repeating Fast radio bursts (FRBs), those precisely localized offer the best opportunities to explore their multi-wavelength counterparts, which are key to uncovering their origins. Here we report our X-ray, ultraviolet (UV), and optical observations with the $Swift$ satellite of the repeating \FRB, in coordination with simultaneous radio observations with the Five-hundred-meter Aperture Spherical radio Telescope (FAST). Our aim was to detect potentially associated multi-wavelength bursts and identify any persistent multi-wavelength counterpart to the associated persistent radio source (PRS). While a total of 10 radio bursts were detected by FAST during the $Swift$ observations, we detected no X-ray, UV, or optical bursts in accompany with the radio bursts. We obtained the energy upper limits ($3\sigma$) on any multi-wavelength bursts as follows: $5.03 \times 10^{47}$ erg in the hard X-ray band (15--150 keV), $7.98 \times 10^{45}$ erg in the soft X-ray band (0.3--10 keV), and $4.51 \times 10^{44}$ erg in the U band (3465\AA), respectively. The energy ratio between soft X-ray (0.3--10 keV) and radio emission of the bursts is constrained as  $<6\times10^{7}$, and the ratio between optical (U band) and radio as $<1.19\times10^{6}$. We detect no multi-wavelength counterpart to the PRS. The 3$\sigma$ luminosity upper limits are 1.04$\times10^{47}$ (15--150 keV), 8.81$\times10^{42}$ (0.3--10 keV), 9.26$\times10^{42}$ (UVW1), and 2.54$\times10^{42}$ erg s$^{-1}$ (U), respectively. We show that the PRS is much more radio loud than representative pulsar wind nebulae, supernova remnants, extended jets of Galactic X-ray binaries, and ultraluminous X-ray sources, suggestive of boosted radio emission of the PRS.   
\end{abstract}



\section{Introduction} \label{sec:intro}

% \textcolor{red}{ To add more: science for Swift vs. FAST observations}

Fast radio bursts (FRBs) are radio bursts with a typical duration on the millisecond timescale. They have been observed with dispersion measure (DM) larger than that of the Galactic values, indicating their cosmological origins \citep[see][for a review]{zhang_physics_2023}. Due to their short duration, localization of the FRBs has been challenging in the radio band, with only tens of FRBs precisely localized among a sample of around 1000 so far. The search for potential persistent and transient multi-wavelength counterparts, through ground or space observations, largely depends on their precise localization to the arc-second scale or better. 

The observed FRB population includes repeating FRBs and non-repeating FRBs. The relationship between these two potential sub-categories and whether they originate from distinct sources remains undetermined. Repeating FRBs, particularly those in their active phase, are prime targets to perform multi-wavelength campaigns aimed at identifying potential burst counterparts across different wavelengths on timescales as short as milliseconds. Despite extensive efforts to search for multi-wavelength counterparts in the optical, X-ray and gamma ray bands in the past\citep[e.g. ][]{Scholz2017,sun_search_2019,Guidorzi_2020,Guidorzi2020,Pilia2020,Tavani2020,laha_simultaneous_2022,Hiramatsu2023,Principe2023,Trudu2023A,Kilpatrick2024}, no positive detection had been made by the time when this paper was submitted. 

Among the entire sample, FRB~20121102A is the first known FRB associated with a persistent radio source (PRS) \citep{chatterjee_direct_2017}. Searches for potential multi-wavelength counterparts of the PRS has also yielded non-detections \citep[e.g.][]{chatterjee_direct_2017,Scholz2017,chen_2023,Eftekhari2023}. The observations of the host galaxy of FRB 20121102A were used to set an upper limit on the optical and infrared counterparts of the PRS \citep{Bassa2017,chatterjee_direct_2017,tendulkar_host_2017}.

The Five-hundred-meter Aperture Spherical radio Telescope \citep[FAST; ][]{nan_five-hundred_2011} is equipped with 19-beams receiver to enlarge its field of view in L-band. \FRB~was discovered from archived data in Commensal Radio Astronomy FAST Survey \citep[CRAFTS; ][]{li18_crafts, li_preface_2019}. It is an active repeating FRB source from which four bursts were detected in the drift scan survey in 2019. Three bursts were detected in the same beam during the drift scan, and the fourth was followed up in another beam. By checking the beam pointings at the time of events, FAST observations localized the FRB to a position accurate to a few arc-minutes. Through the VLA DDT/20A-557 program, we localized the bursts to the sub-arcsecond accuracy with the discovery of a PRS in spatial association with the bursts and determine the redshift of the \FRB~as 0.241 after the identification of its host galaxy \citep{niu_repeating_2022}. With the refined burst and PRS positions, we are able to search for multi-wavelength counterparts for the radio bursts and the PRS, respectively.

\FRB~has been persistently active ever since its discovery and remains the only such repeater without any extended dormant periods (e.g. longer than a few days). This makes it an ideal target for follow-up with multi-band observations. It also produced plenty of bursts with a high signal-to-noise ratio (S/N $\sim$ a few tens) as seen with FAST in our campaign. Therefore, multi-wavelength observations conducted simultaneously or quasi-simultaneously with FAST observations could yield valuable upper limits or even potential detection of multi-wavelength bursts in temporal association with these radio bursts. Such results of multi-wavelength bursts would provide crucial information about the central engine of the \FRB~\citep[e.g.][]{nicastro_multiwavelength_2021,zhang_physics_2023}, thereby revealing the nature of the repeating FRBs or the FRBs as a whole. 

On the other hand, the nature of the PRS and its relation to the FRB bursting source remain unclear. The multi-wavelength properties of the PRS can provide a critical and complementary test for current FRB models due to its physical and positional association with the FRB source. As one of the known FRBs associated with a PRS, detecting a potential multi-wavelength counterpart of the PRS would be a major step towards the understanding of the origin of PRS and its physical relation to the FRB source.

The $Swift$ mission can be flexibly scheduled and is able to quickly respond for target-of-opportunity (ToO) observations of transient events in hard X-ray, soft X-ray, and ultraviolet (UV)/optical bands \citep{gehrels_2004}. It consists of three major instruments: the Burst Alert Telescope (BAT), the X-Ray Telescope (XRT) and
the Ultraviolet/Optical Telescope (UVOT). BAT covers the hard X-ray band ($\sim$ 15--150 keV) and is capable of searching for any bright hard X-ray bursts or persistent sources if the target is in its large field-of-view. XRT is a focusing X-ray telescope that covers the soft X-ray band (0.3--10 keV). It has a time resolution of 2.51 seconds in photon-counting (PC) mode and 1.8 milliseconds in windowed-timing (WT) mode, allowing it to detect potential soft X-ray bursts or a persistent X-ray counterpart. UVOT is equipped with six filters in the UV/optical band and can collect event mode data with high time resolution ($\sim$12 milliseconds), enabling it to detect potential UV/optical bursts and also any persistent counterparts. 

Here we report our simultaneous $Swift$ and FAST observations of the \FRB~performed in May and August of 2020. The PRS position of \FRB~is coincident with the burst within the uncertainty at coordinates 16:02:04.266, -11:17:17.33, obtained from the VLA observation \citep{niu_repeating_2022,zhang2023}. We used these coordinates in the following $Swift$ data analysis to search for potential burst counterpart on timescale of seconds to milliseconds, and to search for the PRS counterpart with long exposures lasting thousands to tens of thousands of seconds.

\section{Observations} \label{sec:obs}
We requested two sets of simultaneous $Swift$ and FAST Target-of-Opportunity (ToO) observations of \FRB~in May and August of 2020. The first $Swift$/FAST campaign in May 2020 aimed to detect potential X-ray bursts in both soft X-ray band (XRT) and hard X-ray band (BAT) in association with the radio bursts detected by a single FAST observation. Additionally, we performed an initial investigation of the field in UV wavelengths using $Swift$/UVOT and searched for potential UV bursts. The second $Swift$/FAST campaign in August 2020 was primarily aimed at fast optical photometric observations performed simultaneously with FAST observations after our target was confirmed active in the radio band and a sub-arcsecond localization was achieved by the VLA. 

The corresponding FAST observations were all conducted in tracking mode. The May 2020 campaign targeted the position determined by previous FAST drift scan and tracking observations. The August 2020 campaign pointed at the position obtained by our preliminary localizations of the radio bursts and the PRS measured with the VLA at the time of the observations. 

\subsection{FAST observations} \label{sec:fast}
%\textcolor{red}{The description of observations is a bit confusing here. Need to specify the year when necessary.}
%{\bf \red Chenhui: please add an introduction of FAST observations}
The FAST observations coeval with the $Swift$ observations were obtained through Director's Discretionary Time (DDT2020$\_$3) and the FAST key science project. Based on the initial coarse localization of \FRB~from the analysis of partially overlapping drift scans as part of the CRAFTS design \citep{li18_crafts},  FAST observations of \FRB~started on April 25, 2020, during which multiple bursts were successfully detected. On May 22, in the FAST observation conducted jointly with $Swift$, 13 bursts were observed, further indicating that this source was active. 

In late July, after our VLA DDT program had successfully localized the bursts to arc-second accuracy, we requested an additional simultaneous campaign on \FRB~with $Swift$ and FAST, utilizing the precise localization. This campaign lasted from August 4 to August 16, 2020, during which a total of 7 $Swift$ observations were carried out, one every two days (\autoref{tab:obs_swift}). During this period, 6 FAST observations were performed. FAST could track \FRB~for $\sim$1.7 hours each day due to the constraint by the observation window. In total, 40 radio bursts were detected at L band by FAST during this campaign \citep{niu_repeating_2022}, 10 of which are covered by the time windows of the $Swift$ observations (\autoref{tab:bursttab}).

\subsection{{\it Swift} observations} 
\label{sec:swift}

We requested a single $Swift$ ToO observation on 2020 May 22 and seven additional ones during the first half of 2020 August \autoref{tab:obs_swift}, which were coordinated with FAST observations. These observations allowed us to search for potential simultaneous optical bursts down to a timescale as short as $\sim$ 12 ms (event mode of U band), and for potential simultaneous X-ray bursts down to a timescale of 2.51 seconds (PC mode of XRT). Since no BAT event data were taken during those targeted $Swift$ observations, we retrieved the BAT event data with exposures longer than 100 seconds and within 30$^{\circ}$ of the \FRB~position during the period of the FAST observations \citep[MJD 58962--59112;][]{niu_repeating_2022} to search for potential hard X-ray counterpart of the radio bursts. The observation information is listed in \autoref{tab:bat}.

In the $Swift$ archive, we identified a previous XRT observation from 2016 that covered the field of \FRB. This observation, triggered by a non-GRB event, was pointed approximately 5 arc-minutes away from \FRB. We have included this observation in our analysis to constrain the X-ray counterpart of the PRS (see \autoref{tab:obs_swift}). We utilize all the $Swift$ observations in \autoref{tab:obs_swift} to the search for potential PRS counterparts in hard X-ray (survey data of BAT), soft X-ray (XRT), optical (U band) and ultraviolet (UVW1 band).

\section{Data analysis and results} \label{sec:analysis}

\begin{table*}
%\caption{The information of the Swift observations and upper limits}
\caption{The information of the {\it Swift} observations and the upper limits
\label{tab:obs_swift}}
\centering
%\scriptsize
\begin{tabular}{ccccccc}
\hline

obsID & start time & exposure & filter  &  AB Mag  & Flux & XRT rate \\
      &  (UTC)    &   (s)      &           &      & ($\mu$jy)  & ($10^{-3}$ c/s)  \\
\hline
00034397001 & 2016-02-27 22:58:04 &  744    & U &  $>21.73$  & $<7.50$ & $<12.39$ \\
00013503001 & 2020-05-22 17:09:35 &  1396  &  UVW1  & $>22.24$ & $<4.54$& $<6.31$ \\
00013503002 & 2020-08-04 11:29:35 & 1378  & U     &$>22.22$  & $<4.77$ & $<6.27$ \\
00013503003 & 2020-08-06 11:12:35  & 1669  &  U    & $>22.33$  & $<4.33$ & $<5.22$ \\
00013503004 & 2020-08-08 10:57:34 &  1558  &  U    & $>22.27$   &$<4.56$& $<7.70$   \\
00013503005 & 2020-08-10 10:53:35 &  1411  &  U    & $>22.15$   & $<5.10$ & $<6.37$ \\
00013503006 & 2020-08-12 10:39:35 &  1557  &  U   & $>22.26$   & $<4.60$ & $<6.17$  \\
00013503007 & 2020-08-14 10:28:36 &  1398  &  U   & $>22.19$     &$<4.89$ & $<5.80$ \\
00013503008 & 2020-08-16 11:42:35 &  1368  &  U  & $>22.13$    &$<5.19$ & $<6.51$  \\
\hline
\end{tabular}   
 
\tablenotetext{}{
Note: All of them correspond to $3\sigma$ upper limits.}
\end{table*}

\subsection{upper limits on the persistent emission }
\label{sec:up_prs}
 
Assuming a power law X-ray spectrum with a photon index of 2 and a Galactic hydrogen column density of  2.6$\times10^{21}$ cm$^{-2}$\footnote{\url{https://www.swift.ac.uk/analysis/nhtot/index.php}}, an XRT rate of 1 counts s$^{-1}$ corresponds to unabsorbed X-ray flux of 5.77$\times10^{-11}$ erg cm$^{-2}$ s$^{-1}$ in the 0.3--10 keV band. Similarly, a BAT rate of 1 counts s$^{-1}$ cm$^{-2}$ corresponds to unabsorbed X-ray flux of 8.33$\times 10^{-8}$ erg cm$^{-2}$ s$^{-1}$ in the 15--150 keV band, according to the conversion from WebPIMMS\footnote{\url{https://heasarc.gsfc.nasa.gov/Tools/w3pimms help.html}}. These fluxes correspond to 9.94$\times 10^{45}$ erg s$^{-1}$ (0.3--10 keV) and 1.48$\times10^{49}$ erg s$^{-1}$ (15--150 keV) for the distance of 1218 Mpc of \FRB~\citep{niu_repeating_2022}. We can use this conversion to roughly estimate the X-ray flux in the following analysis.

For each Swift/XRT observation, we determined the 3$\sigma$ upper limit (see \autoref{tab:obs_swift}) using the online Swift/XRT products generator\footnote{\url{https://www.swift.ac.uk/user\_objects/}} \citep{evans_online_2007}, as there were no positive detection in individual observation at the PRS position of \FRB. We then stacked all the nine XRT observations and obtained the X-ray count rate upper limit of 8.6 $\times 10^{-4}$ c/s at the PRS position. Using the aforementioned flux conversion, the upper limit on the 0.3-10 keV X-ray flux of the potential counterpart of the PRS is 4.96 $\times 10^{-14}$ erg cm$^{-2}$ s$^{-1}$, corresponding to an X-ray luminosity of 8.81$\times10^{42}$ erg s$^{-1}$. 

For each $Swift$/UVOT observation, photometric analysis was performed using the task \texttt{uvotsource} with a recommended aperture radius of 3$\arcsec$\footnote{\url{https://swift.gsfc.nasa.gov/analysis/threads/uvot_thread_aperture.html}}. All the photometric results of each observation are listed in \autoref{tab:obs_swift}. None of the observations yielded a positive detection of the persistent source at the PRS position of the \FRB. To determine the limiting magnitude of the optical counterpart of the PRS, we stacked the images from the eight observations performed with the U filter (see \autoref{tab:obs_swift}). We achieved a 3$\sigma$ upper limit as 23.36 (AB magnitude) in the $U$-band for the persistent source at the \FRB~position, corresponding to a luminosity of 2.54 $\times10^{42}$ erg s$^{-1}$. There was only one observation performed with the UVW1 filter; the upper limit of the PRS is 22.25 (AB magnitude), corresponding to 9.26$\times10^{42}$ erg s$^{-1}$.

The BAT survey data were processed with the \texttt{batsurvey} script. Subsequently, we employed \texttt{batcelldetect} to determine the SNR at the PRS position in BAT sky images. The highest SNR obtained was 2.3$\sigma$, which is significantly below the recommended detection threshold of $5\sigma$. The upper limit of the hard X-ray counterpart of the PRS can be estimated from background level. The $3\sigma$ upper limit is about 0.007 counts s$^{-1}$ cm$^{-2}$ (15--150 keV) with exposure time 1677 seconds. This corresponds to the X-ray flux of $5.83\times 10^{-10}$ erg s$^{-1}$ cm$^{-2}$, assuming a power-law spectrum with index of $2$, and an X-ray luminosity of 1.04$\times10^{47}$ erg s$^{-1}$.

\begin{figure*}
    \centering
%    \begin{subfigure}{0.49\textwidth}
    \includegraphics[width=0.45\linewidth]{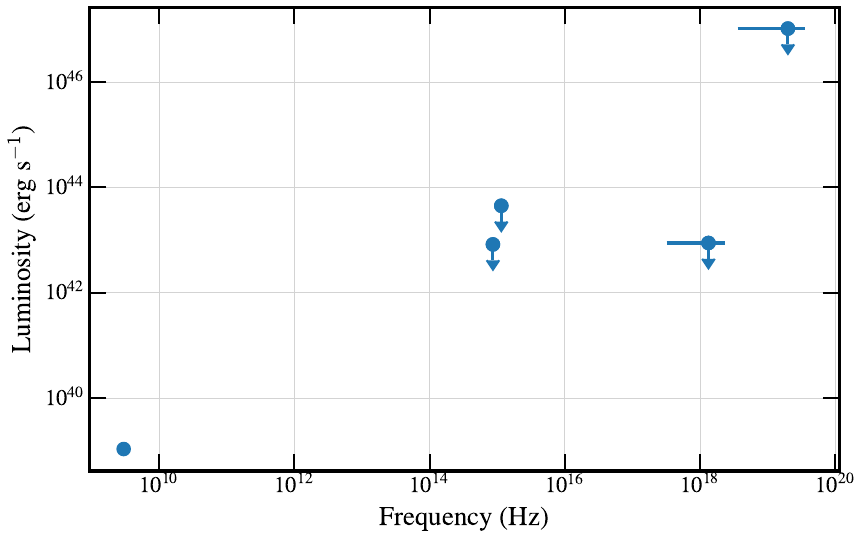}
%   \end{subfigure}
%   \begin{subfigure}{0.49\textwidth}
    \includegraphics[width=0.45\linewidth]{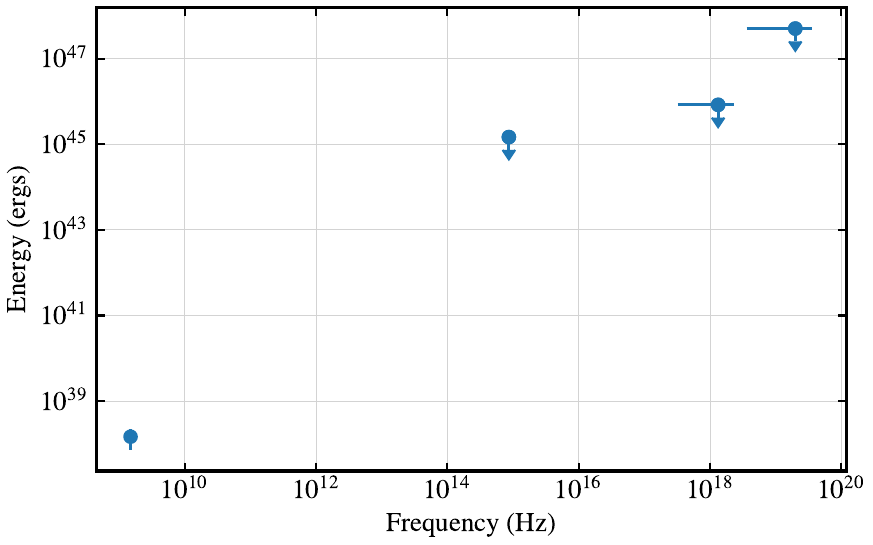}
 %   \end{subfigure}
    \caption{Left: The broad band SED of the PRS of the \FRB. The radio luminosity is taken from \cite{niu_repeating_2022}; the upper limits in wavelengths other than the radio band are obtained in this work. Right: The broad band SED of the burst emission of \FRB. The radio energy is averaged over the burst sample listed in \cite{niu_repeating_2022}; the upper limits in wavelengths other than the radio band are constrained and obtained in this work.}
    \label{fig:SED}
\end{figure*}

We then plotted the broad band spectral energy distribution (SED, \autoref{fig:SED}). The radio flux of PRS is used as $202\pm8~\mu$Jy \citep{niu_repeating_2022}. For the calculation of U and UVW1 fluxes, we applied an extinction of $E(B-V)=0.26$ \citep{niu_repeating_2022}.

\subsection{upper limits on the burst emission}
\label{sec:up_burst}

We also estimate the upper limits on the simultaneous X-ray and optical fluxes of the radio burst detected by FAST. The dispersion delay corresponding to infinite frequency is calculated by using the DM in \autoref{tab:bursttab}. Subsequently, we correct the burst time to search for the potential simultaneous bursts at X-ray and optical bands. 

Since no BAT event data were obtained in the observations listed in \autoref{tab:obs_swift}, we used the observations listed in \autoref{tab:bat} to search for potential hard X-ray bursts, as \FRB~remained active during this period. We extracted the light curves (15--150 keV) at the burst position using the tool \texttt{batbinevt} with a time resolution of 10 ms. This time resolution was chosen because it is comparable to the typical duration of the radio bursts observed with FAST and meets the photon statistics required for the burst search. Subsequently, we searched for coincident X-ray bursts in the 15--150 keV band based on the dispersion delay corrected burst arrival times. Unfortunately, none of the radio bursts detected by FAST were covered by the $Swift$/BAT event data within a span of 1000 seconds.

We then used the standard deviation of the count rate on 10 ms timescale to estimate the 3$\sigma$ upper limit on potential X-ray bursts as 3.40 counts s$^{-1}$ cm$^{-2}$. This corresponds to an X-ray flux (15--150~keV) of 2.83 $\times10^{-7}$ erg s$^{-1}$ cm$^{-2}$, assuming a power law index of 2, and a luminosity of 5.03 $\times 10^{49}$ erg s$^{-1}$. The burst energy can be estimated according to $L\times\delta t$, where $L$ is the upper limit of the luminosity and $\delta t$ is the time resolution. The upper limit of the luminosity in the non-detection case is determined by the instrumental sensitivity, which depends on the chosen time resolution. Therefore, we opted to use energy as the upper limit measurement of the burst emission rather than luminosity, as it is independent of the choice of a certain time resolution, and ensures a homogeneous comparison across the outcomes of instruments with different time resolutions. The upper limit of the burst energy in the 15--150 keV band is $\sim$ 5.03 $\times 10^{47}$ erg.

The $Swift$/XRT observations were performed in the PC mode.
In order to estimate the soft X-ray upper limit for each radio burst detected by FAST, we used the online tool \citep{evans_online_2007} to generate the light curve of each observation with a time resolution of 2.51 seconds. A total of 10 radio bursts were simultaneously covered by the $Swift$/XRT observations; with upper limits on the burst emission ranging from 3.20 to 3.63 c/s. The corresponding upper limits on the X-ray luminosity and energy are roughly estimated as (3.28--3.72)
$\times10^{46}$ erg s$^{-1}$, and (8.22--9.34)
$\times10^{46}$ erg. We also searched for potential X-ray burst within the 100 s interval around each radio burst. No positive detection is archived, and the upper limit remain within the aforementioned range. It is worth noting that only two photons detected within the PSF ($18\arcsec$) of the burst position during the total exposure of $\sim$12400 seconds. 

We then extracted available $Swift$/XRT images coincident with the radio burst using \texttt{xselect}, with a time resolution of 2.51 seconds. Corresponding exposure maps are also produced. We then stacked all the 10 XRT sky images and exposure maps, and obtained a $3\sigma$ upper limit as 0.32 counts~s$^{-1}$ using the tool \texttt{sosta}. This count rate corresponds to an upper limit on X-ray luminosity of 3.18\ $\times10^{45}$ erg s$^{-1}$, and an upper limit on X-ray energy of 7.98$\times10^{45}$ erg.

We also notice that the DM is very large $\sim$ 1200 pc cm$^{-3}$ for \FRB. According to the correlation between the DM and X-ray column density in \citet{He2013ApJ}, the corresponding column density of \FRB~is about 3.6$\times10^{22}$ cm$^{-2}$. The unabsorbed X-ray flux observed by the XRT in the 0.3--10 keV band is about four times larger than previous estimations, which were based solely on Galactic column density in the source's direction. However, the unabsorbed X-ray flux as seen with BAT in the 15--150 keV range remains unchanged. The origin of DM is still uncertain for these bursts \citep{niu_repeating_2022}, and local contributions are likely significant \citep{2023ApJ...954L...7L}. For the purposes of order-of-magnitude estimation, we will continue to use the X-ray flux upper limits derived from the Galactic column density in the followings.

One of the aims of the August 2020 campaign was to perform fast photometry in the optical band to search for the optical burst counterparts. Seven $Swift$/UVOT observations with U filter (3465\AA) were conducted in event mode. We first used \texttt{uvotscreen} to obtain the corresponding cleaned event data. The astrometry of the event data is then refined by using the method in \citet{oates_2009}. We extracted $U$-band images every 100 s. The position of each photon event was corrected based on the differences obtained from cross-correlating the image with the USNO-B1 catalogue. We then used the \texttt{uvotevtlc} task to generate the corresponding light curve with a time resolution of 12 milliseconds from each event file, applying a circular source region with a radius of 3$\arcsec$. Consequently, the upper limits on the source flux in the time resolution of 12 milliseconds, which were simultaneous with any radio burst detected by FAST, were obtained. The upper limits on the luminosity and burst energy on the timescale of 12 ms for 9 available bursts are the same as $3.76\times 10^{46}$ erg s$^{-1}$ and $4.51\times 10^{44}$ erg. 
We also searched the 100s interval around each radio burst but found no positive detection. The upper limit remains the same.

We then use the burst energy upper limit extracted from the stacked XRT images, BAT event data and U band event data to plot the broad band SED of the burst emission in \autoref{fig:SED}. The average burst energy at 1.5 GHz and E(B-V)=0.258 from \citet{niu_repeating_2022} is adopted .

Using the average radio burst energy in \citet{niu_repeating_2022} and the X-ray burst energy upper limit from stacked XRT images, we derived the upper limit of the X-ray-to-radio energy ratio $E_{\rm X}/E_{\rm R} < 6\times10^{7}$.

\section{Discussion} \label{sec:discussion}

\begin{figure*}
    \centering
    \includegraphics[width=\linewidth]{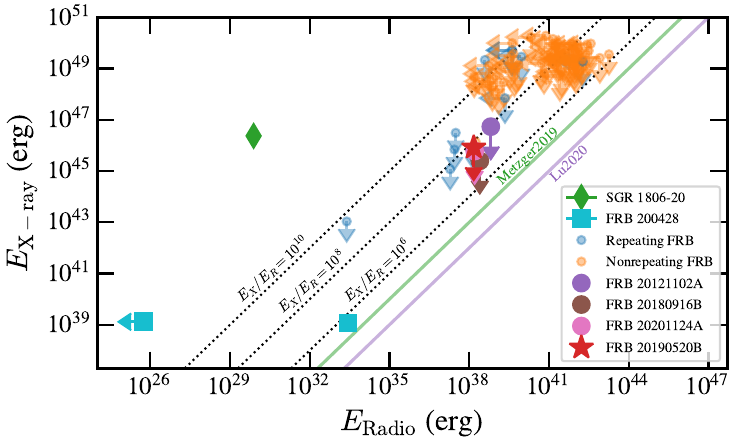}
    \caption{The relationship between X-ray and radio energy of bursts in the FRBs. The data, except for \FRB, are taken from \citet{Scholz2017,Guidorzi_2020,  laha_simultaneous_2022}. The green and purple lines are the model prediction from \citet{metzger19} and \citet{lu20}, respectively. For many FRBs, the distances are constrained as upper limits, so the burst energy in both X-ray and radio bands are also upper limits.}
    \label{fig:E_RX}
\end{figure*}

Repeating FRBs offer a valuable opportunity for burst localization and the search for multi-wavelength counterparts. However, over the past few years, significant challenges have persisted in the efforts towards the detection of FRB counterparts, both from space and ground-based observations. In May and August 2020, we conducted simultaneous $Swift$ pointed observations with FAST tracking observations of the actively repeating fast radio burst \FRB, the second known FRB associated with a compact PRS. Here, we report the results of multi-wavelength counterparts of both the bursts and the PRS from our $Swift$ campaign. Utilizing short and long exposures from the BAT, XRT and UVOT onboard $Swift$, we placed constraints on potential multi-wavelength counterparts to the radio bursts and the PRS, based on precise localization of the source by the VLA in 2020.

\subsection{Burst counterparts}
In our campaign, $Swift$ observations simultaneously covered 10 radio bursts detected by FAST (see \autoref{tab:bursttab}). No X-ray or optical burst events were detected.
We derived upper limits for the burst flux on short timescales in both the soft X-ray (0.3--10 keV, 2.51 seconds), hard X-ray (15--150 keV, 10 ms) and optical (U band, 12 ms) bands (see \autoref{sec:up_burst}). The current measurements are constrained by the sensitivity of the instruments at these short timescales. Since the upper limit on burst luminosity depends on the time resolution used, it is more appropriate to use the upper limit on burst energy to constrain the multi-wavelength emission.

\citet{chen_multiwavelength_2020} estimated the ratio $\eta$, which represents the energy of FRB counterparts at various wavelengths compared to the radio wavelength, based on the non-detection from various surveys and instruments. This ratio is crucial for constraining emission models across multiple wavelengths.
We derive a lower limit of the ratio $\eta=E_{\rm X}/E_{\rm R} \lesssim 6\times10^{7}$ and $\eta=E_{\rm U}/E_{\rm R} \lesssim 10^{6}$. Additionally, using the radio and X-ray burst energy values from an FRB sample reported by \citet{laha_simultaneous_2022} and \citet{Guidorzi_2020}, we re-plotted the relation between $E_{\rm R}$ and $E_{\rm X}$ (\autoref{fig:E_RX}).

The upper limits on the burst fluences in the X-ray band are mainly constrained by the instrument sensitivity at short timescales. Notably, the values of upper limits for $E_{\rm X}$ tend to saturate around ${10}^{50}~{\rm ergs}$. For many FRBs, the burst energy in both X-ray and radio bands are constrained as upper limits, because their distances correspond to upper limits. For comparison, we also marked the upper limits obtained for FRB 20121102A \citep{Scholz2017}, FRB 20180916B \citep{Pilia2020,Trudu2023A} and FRB 20201124A \citep{Piro2021} in \autoref{fig:E_RX}. These FRBs have been active, with extensive multi-wavelength campaigns to search for the burst counterparts. They show comparable upper limits of $E_\mathrm{X}/E_\mathrm{R} \sim 10^{7}-10^{8}$, similar to that of \FRB.

The discovery of Galactic FRB 20200428 \citep[][SGR 1935+2151; ]{bochenek_fast_2020} suggests a potential magnetar origin for FRBs, though the energy budget remains a major concern for magnetar models \citep{li_bimodal_2021}. FRB 20200428 is much weaker than most FRBs, by several orders of magnitude. On the other hand, thousands of bursts from FRB 20121102A were observed within a time span of 47 days, adding up to a substantial fraction of total available magnetic energy from a magnetar \citep{li_bimodal_2021}. The prolonged burst activities of \FRB~poses even greater challenges for the magnetar models. 

FRB emission generally fall into two categories in the magnetar-based models: those invoking emission inside the magnetosphere of the magnetar \citep[e.g.][]{kumar17,yangzhang18,lu20} and those invoking relativistic shocks outside the magnetoshere 
 \citep[e.g.][]{lyubarsky14,metzger19,beloborodov20}. The latter predicts a relatively large $E_{\rm X}/E_{\rm R}$ \citep{sironi21}, while the former allows for a smaller ratio \citep{zhang20}. 
 Our constraints on \FRB~exceed the prediction of both models \citep[\autoref{fig:E_RX}, see also][]{chen_multiwavelength_2020}. More sensitive X-ray/optical telescopes are required for current FRB samples, or only very nearby extragalactic X-ray/optical bursts will be detectable if these model predictions hold. 

\subsection{PRS counterpart}
\begin{figure*}
    \centering
    \includegraphics[width=\linewidth]{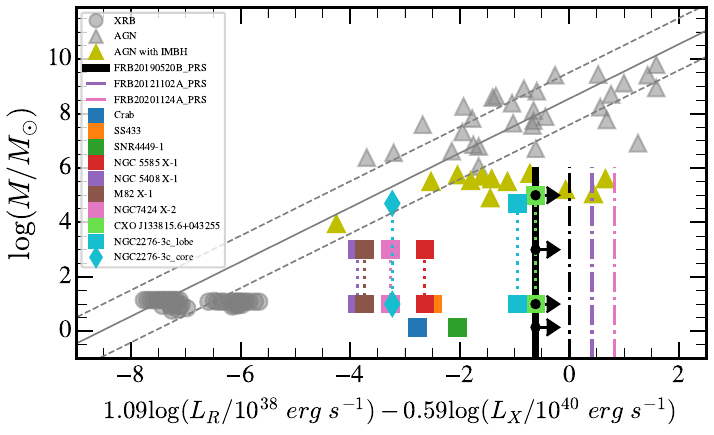}
    \caption{The gray data points, solid lines, and dashed lines represent the data for XRBs, AGNs, and best-fitting results for the fundamental plane of accretion and jet activity of different scales of BHs \citep{gultekin_2019}. Different kinds of systems are also plotted in this diagram, including PWN, SNR, XRB nebula and ULXs. The solid black line represents the lower limit of the PRS of \FRB~derived in this work, and the black dots with arrows are assumed BH masses of $10^{5}$, $10^{3}$, $10$ and $1.4$ $M_{\odot}$. The black dash-dotted line is the lower limit of the PRS of \FRB~calculated from the X-ray constraint with $Chandra$. The purple and pink dash-dotted lines represent the lower limit of the PRSs for FRB 20121102A and FRB 20201124A.}
    \label{fig:M_RX}
\end{figure*}

The PRS in association with the \FRB~exhibits peculiar variability, suggesting the possibility of an accreting compact object scenario \citep{zhang2023}. Constraining the multi-wavelength counterpart is crucial for uncovering and understanding the true nature of the PRS. By stacking the XRT observations, we obtained an upper limit on the persistent flux of 8.81$\times 10^{42}$ erg s$^{-1}$. 
Additionally, the upper limits on a potential persistent counterpart were obtained as 2.54$\times 10^{42}$ erg s$^{-1}$ at U band and 9.26$\times 10^{42}$ erg s$^{-1}$ at UVW1 band. 

As the two active FRBs and the best known FRBs associated with a compact PRS, \FRB~and FRB~20121102A share striking similarities in their burst environment. Both are hosted by dwarf galaxies with high star formation rates \citep{chatterjee_direct_2017,niu_repeating_2022}. If the central engine is a magnetar, the PRS could correspond to a pulsar wind nebula (PWN) or a supernova remnant (SNR). However, the radio flux and spectra of the PRSs show variations on timescales as short as days, suggesting that a significant fraction of the instant PRS radio emission may result from accretion onto a compact object \citep{zhang2023,rhodes2023}. Additionally, their extremely large and variable rotation measures imply that they exist within dynamic, highly magnetized environments \citep{michilli_extreme_2018,thomas2022}. 

On the other hand, it has been suggested that binary systems may lie behind FRB sources \citep[e.g. ][]{chen2022ApJ}. The periodic activity of FRB~20121102A \citep{Rajwade2020} and the sign reversal of the rotation measure of \FRB~\citep{thomas2022} have led to models proposing a binary system. \cite{Sridhar2022} have put forward a scenario where a nebula surrounding a hyper-accretion X-ray binary (similar to an ultraluminous X-ray source; ULX) might correspond to the PRS of the FRB. In this scenario, there should be a persistent X-ray counterpart to account for the ULX origin. Many of the radio properties of the PRS of the \FRB~are consistent with the hyper-accretion binary scenario, particularly when the apparent radio variability \citep{zhang2023} is attributed to scintillation \citep{2023ApJ...958L..19B}. Recently, off-nuclear compact radio sources have been detected in some dwarf galaxies \citep{Reines2020ApJ}, likely powered by an accreting intermediate-mass black hole (IMBH, $10^{3}$--$10^{6} M_{\odot}$). These sources exhibit many similarities to the two FRB PRSs, including their radio luminosity, spectra and variability \citep{Eftekhari2020,Law2022ApJ}.

If the source behind the PRS is an accreting compact object, which may include but is not limited to hyper-accretion systems, we can compare our derived properties in the radio and X-ray band with those measurements taken from accreting black holes (BHs) and neutron stars. There is an established, empirical fundamental plane that characterizes accretion and jet activities of BHs, which demonstrates a universal correlation between X-ray luminosity, radio luminosity, and BH mass across a broad range of masses, encompassing both X-ray binaries (XRBs) and AGNs. We have re-plotted this fundamental plane using the data and best-fitting results from \citet{gultekin_2019}. Additionally, we included some samples of the AGNs harboring IMBHs compiled by \citet{Yang2023G}. Some of these samples show obvious deviations from the fundamental plane \citep[\autoref{fig:M_RX}, see also ][]{Gultekin2022M}. 
The boosted radio emission and/or extended jet could cause such deviations \citep{Yang2023G}.

We also plotted some extragalactic ULXs \citep{Laor2008,Mezcua2015,Mezcua2018MNRAS,Soria2021}, which are thought to be IMBH candidates or super-Eddington accreting stellar-mass compact objects. For these ULXs, we plotted both 10 $M_{\odot}$ BH mass and the mass estimations in the literature (or 1000 $M_{\odot}$ if no mass estimate was available). The nature of the radio emission from ULXs is still under debate. Some are thought to originate from jet emission, which tends to be more radio-loud compared to emission from nebulae. High resolution VLBI observations of these sources can resolve the jet structures. For example, both a radio core and large-scale radio lobes have been detected in NGC 2276-3c \citep{Mezcua2015}. The radio core emission from this source aligns with the fundamental plane if the black hole mass is assumed to be approximately $5\times10^{4}~M_{\odot}$. Similarly, two radio lobes have also been observed in the ULX CXO J133815.6+043255 in NGC 5252. The total radio emission from both ULXs is believed to arise from extended jets rather than a compact core jet. 

Since the position of the PRS of \FRB~is off-nuclear \citep{niu_repeating_2022}, its radio emission can not be attributed to the activity of the central super-massive or intermediate-mass BH in the dwarf galaxy. If the PRS is powered by an accreting BH, the BH mass is likely less than $10^{6} M_{\odot}$, as inferred from the fundamental plane (see also \citealt{zhang2023}). So we plotted a vertical line below $10^{6} M_{\odot}$ as the lower limit of the PRS of \FRB~in the diagram of \autoref{fig:M_RX}. We also display the lower limits corresponding to different compact object masses: 1.4, 10, 10$^3$ and 10$^{5}~M_{\odot}$. 

\citet{Eftekhari2023} have given a more stringent upper limit of $6.57\times10^{41}$ erg s$^{-1}$ based on a 14.9 ks $Chandra$ observation. The lower limit range calculated from this constraint is also plotted as a vertical dash-dotted line in black. The lower limit range corresponding to the first detected PRS associated with FRB~20121102A is also plotted \citep{chatterjee_direct_2017,Eftekhari2023}. FRB~ 20201124A is the third FRB found associated with a PRS \citep{Bruni2024,Dong2024}. However the size of its reported PRS is significantly more extended than the other two \citep{chatterjee_direct_2017,Bhandari2023,Bruni2024}, which suggests that it might be a PRS with significant differences from the two more compact PRSs. We used the reported flux of the compact radio emission \citep{Bruni2024} and the X-ray upper limit \citep{Piro2021} to calculate the lower limit range. The three PRSs all reside in the lower right region below the fundamental plane of the accreting BHs in \autoref{fig:M_RX}.

To illustrate additional representative systems that have been considered as potential candidates for PRSs, we have included the data from the Crab nebula \citep{Lyutikov2019}, an extragalactic SNR \citep{Mezcua_2013}, and a nebula of the Galactic XRB SS 433 \citep{Wolter2015}. Obviously, these sources do not follow the fundamental plane, since their radio emission is not powered by compact jets as in accreting black hole systems. 

We summarize the insights from the diagram of \autoref{fig:M_RX}. The PRS of \FRB~is more radio-loud than the brightest SNR, Crab nebula, SS433 and several ULXs. Notably, the two ULXs with extended radio jets (namely CXO J133815.6+043255 and NGC 2276-3c) and a few AGNs harboring IMBHs are positioned close to the lower limit of the three FRBs (see \autoref{fig:M_RX}). The radio spectral index of the PRS of \FRB~is -0.40$\pm$0.06 \citep{zhang2023}, which is also similar to the extended jet of the two ULXs \citep[-0.5$\pm$0.2 for NGC 2276-3c and $-0.66 \pm$ 0.02 for CXO J133815.6+043255; ][]{Mezcua2015,Smith2023}. However, the radio luminosity of the PRS of \FRB~($\sim10^{39}$ erg s$^{-1}$) is roughly two orders of magnitude larger than that of the radio jets of the two ULXs ($\sim10^{37}$ erg s$^{-1}$). This suggests that the PRS of \FRB~must either possess a substantially more luminous extended radio jet or the observed radio jet emission is significantly beamed and boosted if corresponding source is a ULX of a similar type, as discussed for a potential PRS origin of accreting compact objects with an uncertain range of masses \citep{zhang2023}. It is worth noting that the rare association of FRBs with a PRS within the entire FRB population (only a few cases out of a thousand or more FRBs), can be straightforwardly explained if the radio emission of the PRS of \FRB~is beamed in a small solid angle, on the order of 1/1000 of 4$\pi$, and boosted by relativistic effects.

\begin{acknowledgments}
We would like to thank the anonymous referee for stimulating and helpful comments and suggestions. WY would like to thank the $Swift$ PI, Brad Cenko (and his designate) for approving and scheduling our Swift observations. We appreciate the $Swift$ team for helping with quick data access. We would like to thank the FAST TAC to approve our DDT observations and the FRB Key Science Project for the arrangement of half of the FAST observations that are reported in this paper. WY, ZY and DL would like to acknowledge support by the National Natural Science Foundation of China (grant number 11333005, U1838203, U1938114, 11988101, 12361131579, 12373049 and 12373050). ZY was also supported in part by the Youth Innovation Promotion Association of Chinese Academy of Sciences. DL is a New Cornerstone Investigator.
KLP acknowledges funding from the UK Space Agency.

\end{acknowledgments}
 

\vspace{5mm}
\facilities{Swift, FAST}


\software{astropy \citep{astropy2018}, matplotlib, proplot\citep{luke_2021},HEASoft}

\bibliography{references}{}
\bibliographystyle{aasjournal}

\clearpage

\appendix
%\section{}

\begin{table}%
\caption{ The observations of $Swift$/BAT for the event data analysis
\label{tab:bat}}
%\scriptsize
\centering
\begin{tabular}  {cc c }
\hline%
\hline%
%Burst\ ID & Burst time & DM & Pulse Width & Bandwidth & Scatter tail & Fluence & Energy\\%
obsID & start time  & exposure   \\
      & (UTC)        & (seconds)  \\
00968731000& 2020-04-30 12:53:02& 120\\
00972010000& 2020-05-13 13:23:32& 272\\
00972011000& 2020-05-13 13:33:27& 1023\\
00972030000& 2020-05-13 14:51:48& 1142\\
00973140000& 2020-05-19 11:16:44& 1202\\
00088915014& 2020-05-20 14:10:19& 200 \\
00973140006& 2020-05-22 15:36:09& 200 \\
00974942000& 2020-05-29 00:52:56& 1122 \\
00975895000& 2020-06-03 19:25:16& 1202\\
00089076001& 2020-07-14 05:55:33& 121 \\
00095656005& 2020-07-18 08:24:09& 200 \\
00013597029& 2020-08-23 11:13:08& 200 \\
00095660154& 2020-09-16 06:01:40& 200 \\
00996184000& 2020-09-17 03:48:49& 1202 \\
00996184003& 2020-09-19 14:58:12& 200 \\
%          & (MJD)      & ($\dmu$)   & (ms) & (GHz)  & (mJy/ms) & ($\times10^{37}$erg) \\
\hline%
%7 & 58991.684351079 & 1200.4(20.8) & 17.5(2.0) & 1300-1450 & 0.0(0.0) & 0.0(0.0
%) \\

\hline
\end{tabular}  
\end{table}

\begin{table*}%
\caption{ The properties of the detected FRB bursts during the $Swift$ observations.
}
\label{tab:bursttab}
%\scriptsize
\centering
\begin{tabular}  {llllll}
\hline%
\hline%
%Burst\ ID & Burst time & DM & Pulse Width & Bandwidth & Scatter tail & Fluence & Energy\\%
 Burst time$^{a}$ & DM & Pulse width  & Energy  & XRT time & U time \\
  (MJD)  & (pc cm$^{-3}$) & (ms)  & ($\times10^{37}$erg)  & (MJD)  & (MJD)\\
\hline%
58991.7289360891 & 1214.2 & 30.2 & 16.8  & 58991.728923501 & \nodata \\
59067.484346541 & 1202.8 & 14.9 & 15.9  & 59067.484320859 & 59067.484320927 \\
59067.4843471197 & 1202.8 & 33.1 & 17.8 & 59067.484320859 & 59067.484321482\\
59071.470504066 & 1200.0 & 5.2 & 9.6  &59071.470485296 & 59071.470478477\\
59071.4705042975 & 1200.0 & 7.2 & 8.1 &59071.470485296 &  59071.470478755\\
59075.4527133255 & 1210.6 & 24.9 & 38.0  &59075.452698807  &   59075.452687499\\
59075.453222841 & 1210.6 & 22.0 & 25.5 &59075.453192673   & 59075.453196944  \\
59077.496363178 & 1218.6 & 15.3 & 16.8  &59077.496344162  & 59077.496337235    \\
59077.4965143868 & 1218.6 & 13.6 & 10.9  &59077.496489417  &   59077.496488346  \\
59077.497517645 & 1218.6 & 4.6 & 11.4    &59077.497477148  &  59077.497491680 \\

\hline
\end{tabular}

\footnotesize
\begin{itemize}
\item [$^a$] The burst time is at Coordinated Universal Time (UTC), which is referenced to 1.5 GHz.
 
\end{itemize}
\end{table*}

\end{document}